\documentclass[11pt,fleqn]{article}

\usepackage{amsfonts}
\usepackage{amsmath}
\usepackage{amssymb}
\usepackage{mathtools}
\usepackage{mathrsfs}
\usepackage{enumerate}
\usepackage{bbm}
\usepackage{graphicx,epsfig,amsthm,color}
\usepackage{pstricks}
\usepackage{subfig}
\usepackage{graphicx}
\usepackage{units}

\usepackage{caption}
\usepackage{booktabs}
\usepackage{textcomp}
\usepackage{array}
\usepackage{cite}
\usepackage{cancel}

\date{\today}

\newcolumntype{z}[1]{>{\RaggedRight\hspace{0pt}}p{#1}}
\newcolumntype{w}[1]{>{\RaggedRight\hspace{0pt}}p{#1}}
\newcolumntype{v}[1]{>{\Centering\hspace{0pt}}p{#1}}

\def\be{\begin{equation}}
\def\ee{\end{equation}}
\def\bea{\begin{eqnarray}}
\def\eea{\end{eqnarray}}

\def\be{\begin{equation}}
\def\ee{\end{equation}}
\def\bea{\begin{eqnarray}}
\def\eea{\end{eqnarray}}

\def\erp2{{\rm e}^{2\rho}}
\def\erm2{{\rm e}^{-2\rho}}
\def\er4{{\rm e}^{4\rho}}

\def\be{\begin{equation}}
\def\ee{\end{equation}}
\def\bea{\begin{eqnarray}}
\def\eea{\end{eqnarray}}

\def\m0{m_{\nu_{0,i}}}

\def\T0{T_{\nu_0}}

\newcommand{\half}{\frac{1}{2}}

\newcommand{\beqa}{\begin{eqnarray}}
\newcommand{\eeqa}{\end{eqnarray}}
\newcommand{\bpr}{\begin{problem}}
\newcommand{\epr}{\end{problem}}
\newcommand{\bcent}{\begin{center}}
\newcommand{\ecent}{\end{center}}
\newcommand{\bfig}{\begin{figure}}
\newcommand{\efig}{\end{figure}}
\newcommand{\bpc}{\begin{picture}}
\newcommand{\epc}{\end{picture}}

\renewcommand{\and}{A_{0}^{\nu ,D}(s)}

\newcommand{\bee}{\begin{equation}}

\def\beq{\begin{eqnarray}}
\def\eeq{\end{eqnarray}}

\newcommand{\bright}{\begin{flushright}}
\newcommand{\eright}{\end{flushright}}
\newcommand{\bminip}{\begin{minipage}}
\newcommand{\eminip}{\end{minipage}}

\DeclareCaptionLabelSeparator{mysep}{\hspace{3pt}:\hspace{3pt}}
\DeclareCaptionLabelFormat{mypiccap}{Fig.\hspace{3pt}{#2}}
\DeclareCaptionLabelFormat{mytabcap}{Tab.\hspace{3pt}{#2}}

\begin{document}

\date{}
\title{
\vskip 2cm {\bf\huge Chameleon fields and solar physics}\\[0.8cm]}

\author{{\sc\normalsize Andrea Zanzi$^1$\footnote{E-mail: andrea.zanzi@unife.it} and Barbara Ricci$^{1,2}$\footnote{E-mail: ricci@fe.infn.it}\!
\!}\\[1cm]
{\normalsize $^1$Dipartimento di Fisica e Scienze della Terra, Universit\'a di Ferrara - Italy}\\
{\normalsize $^2$ Istituto Nazionale di Fisica Nucleare, Sezione di Ferrara - Italy}\\[1cm]
}
\maketitle \thispagestyle{empty}

\begin{abstract}
{In this article we discuss some aspects of solar physics from the standpoint of the so-called chameleon fields (i.e. quantum fields, typically scalar, where the mass is an increasing function of the matter density of the environment). Firstly, we analyze the effects of a chameleon-induced deviation from standard gravity just below the surface of the Sun. In particular, we develop solar models which take into account the presence of the chameleon and we show that they are inconsistent with the helioseismic data. This inconsistency presents itself not only with the typical chameleon set-up discussed in the literature (where the mass scale of the potential is fine-tuned to the meV), but also if we remove the fine-tuning on the scale of the potential. However, if we modify standard gravity only in a shell of thickness $10^{-6} R_{\odot}$ just below the solar surface, the model is basically indistinguishable from a Standard Solar Model. Secondly, we point out that, in a model recently considered in the literature (we call this model ''Modified Fujii's Model''), a conceivable interpretation of the solar oscillations is given by quantum vacuum fluctuations of a chameleon.}
\end{abstract}
%PACS numbers: 04.60.Cf, 98.80.-k, 95.36.+x

%\clearpage

%\tableofcontents
\newpage

%%%%%%%%%%%%%%%%%%%%%%%%%%%%%%%%%%%%%%%%%%%%%%%%%%%%%%%%%%%%%%%%%%%%%%%%%%%%%%%%%%%%%%%%%%%%%%%%%%%%%%%%%%%%%%%%%%%%%
% Section INTRODUCTION
%%%%%%%%%%%%%%%%%%%%%%%%%%%%%%%%%%%%%%%%%%%%%%%%%%%%%%%%%%%%%%%%%%%%%%%%%%%%%%%%%%%%%%%%%%%%%%%%%%%%%%%%%%%%%%%%%%%%%
\setcounter{equation}{0}
\section{Introduction}

Helioseismology is one of the possible ways to study solar physics (see e.g. \cite{TurckChieze:2010gc}). In 1916 Plaskett observed fluctuations in the velocity of superficial elements exploiting Doppler effect. In 1962 Leighton detected an oscillation with period $P \simeq 300s$ that was damped after a few minutes.
The first detections were characterized by limited spatial and temporal extensions and, hence, they were typically interpreted as local phenomena in the solar atmosphere. Oscillations in the solar diameter were detected in 1975 by Hill and, for the first time, the global nature of these oscillations was suggested. Today helioseismology is able to provide quantitative information about many solar properties including the speed of sound, chemical composition, the thickness of the convective zone and internal rotation (see e.g. \cite{Basu:2007fp}).

One of the issues that will be investigated in this article is the possibility to exploit solar physics to study the so-called {\it chameleon particles}. These chameleons are quantum fields (typically scalar) with a mass which is an increasing function of the matter density of the environment and they have been originally introduced in \cite{Khoury:2003rn, Khoury:2003aq}. For a general introduction to chameleon fields see \cite{Mota:2006fz, Khoury:2013yy} and references therein. The careful reader may wonder whether the coupling matter-chameleon leads to detectable deviations from standard General Relativity (GR). After all, in the solar system the matter density is small and, hence, the chameleon can be very light (this issue is model dependent). It is common knowledge that it is not possible to introduce in a model a very light scalar degree of freedom with a generic coupling to matter because the light scalar particle would mediate a long-range force and phenomenological constraints must be faced. Naturally, when the matter density is large, the chameleon is heavy and, consequently, the chameleon mediated fifth-force is screened. However, in the solar system the field can be light and another mechanism is required to hide the chameleon in gravity tests: it is called thin-shell mechanism. When this mechanism is operative in the Sun\footnote{The condition that must be satisfied is that $m_c R>>1$, where $m_c$ is the mass of the chameleon inside the object (e.g. the Sun) and $R$ is the radius of the object.}, the chameleon field outside the Sun is determined, to a good approximation, by a thin-shell just below the solar surface. The thin-shell mechanism suppresses the chameleon force and, in this way, phenomenological constraints on deviations from GR can be faced in the solar system. 
Remarkably, the chameleon effect plays also a very important role for constructing viable modified gravities, see for example \cite{Nojiri:2010wj}.

The purposes of this paper are:
\begin{itemize}
\item We analyze the effects of a chameleon-induced deviation from standard gravity in the solar thin-shell. We make explicit numerical simulations exploiting solar models that take into account the presence of the chameleon field. In particular, we discuss, on the one hand, a standard chameleonic scenario where the mass scale of the potential is fine-tuned to the meV Dark Energy scale and, on the other hand, a second chameleonic model where this fine-tuning is removed. We parametrize the presence of the chameleon through a modification of the Newton constant inside the shell. Interestingly, our solar models are {\it inconsistent} with the standard observables of helioseismic observations, namely, a) the thickness of the convective zone; b) the helium abundance on the solar surface; c) the speed of sound as a function of the distance from the center of the Sun. In these models the thickness of the shell is $10\%$ of the solar radius.
\item The thickness of the thin-shell is bounded by Lunar Laser Ranging (see for example \cite{Weltman:2008ll}) to be less than approximately $10^{-10} R_{\odot}$. For this reason we developed also solar models with shells much thinner than $0.1 R_{\odot}$. In particular, we considered also models where the thickness of the shell is $10^{-4} R_{\odot}$ and $2 \cdot 10^{-6} R_{\odot}$. In the case we mentioned last, the shell is the thinnest that can be analyzed by the computer and the results are basically indistinguishable from a Standard Solar Model (SSM). 
\item In order to explore further the connection between chameleon fields and solar physics, we consider a ''Modified Fujii's Model'' (MFM) recently analyzed in \cite{Zanzi:2010rs} and we show that, in this model, a conceivable interpretation of the solar oscillations is given by quantum vacuum fluctuations. The equations of helioseismology are {\it classical} hydrodynamical equations where the attractive contribution of gravity is balanced by pressure (i.e. electromagnetic) force. In this article, however, we suggest to exploit quantum physics to study the oscillations in the Sun. In particular, the solar oscillations are interpreted as quantum vacuum fluctuations of a scalar field (the dilaton) while the electromagnetic pressure force of the classical picture is reinterpreted as a repulsive contribution due to (the gradient of the) quantum vacuum energy. As we will see, the MFM is telling us that helioseismology could be quantum physics.
\end{itemize}

The connection between solar physics and chameleon fields has also been discussed in \cite{Brax:2010xq, Vincent:2012bw, Brax:2011wp, Baker:2012ah}.

As far as the organization of this article is concerned, in section \ref{approx} we discuss some useful formulas for the standard chameleon field; in section \ref{smodels} we build solar models which are properly modified to take into account the presence of a standard chameleon field. In section  \ref{model} we present the lagrangian of the MFM. Some aspects of solar physics are investigated from the standpoint of the MFM in paragraph \ref{helio}, in particular, we discuss quantum helioseismology. In the final part of the paper, we briefly touch upon some concluding remarks.

\setcounter{equation}{0}
\section{The standard chameleon}
\label{approx}

The standard chameleon mechanism is obtained through a ''competition'' between two different contributions to the effective potential: on the one hand, a bare potential (which is not necessarily run-away) and, on the other hand, a matter branch. When the matter density is large, the matter branch is uplifted and a large mass is obtained (i.e. the field is stabilized). On the contrary, on cosmological distances, the matter density is very small, the effective potential is basically indistinguishable from the bare one and the chameleon can be very light. 

In this section we consider one example of chameleonic potential which has been discussed in the literature. In particular, we have the following contributions to the effective potential\footnote{Here and in the following we assume $\hbar =c=1$.}: 1) a run-away bare potential of the form $V=\lambda M^4 (\frac{M}{C})^n$, where $C$ is the chameleon, $n$ is an integer, $M$ is the mass scale of the potential and $\lambda$ is a real number; 2) a matter branch contribution given by an exponential function. In other words, the effective potential is 

\begin{equation}
\label{veff} V_{\text{eff}}(C)\equiv V(C)+\sum_i\rho_i
e^{\left(1-3w_i\right)\beta_i C/M_\text{pl}},
\end{equation}
where $\beta_i$ is a constant parameter (that can be also much larger than one), $w_i$ is the equation of state of the i-th particle specie and $\rho_i$ is the matter density of the i-th particle specie.

The acceleration induced by the chameleon is given by 

\begin{equation}
\frac{\vec{F}_C}{m}=-\frac{\beta_i}{M_\text{pl}}\vec{\nabla}C,
\label{chameleonforce}
\end{equation}. 

The variation of the chameleon field inside the shell can be estimated with 

\begin{equation}
\frac{C}{C_c}=1+\frac{\beta}{M_p} \frac{\rho_c
R_c^2}{C_c} [ \frac{1}{6} (\frac{r}{R_c})^2+\frac{R_c}{3 r}-\frac{1}{2}],
\label{formuletta}
\end{equation} 
where $\rho_c$ is the matter density inside the object, $R$ is the solar radius, $C_c$ is the value of the chameleon field inside the object and the thin-shell is in the region $R_c<r<R$. We can estimate the absolute value of the chameleonic acceleration \ref{chameleonforce} near the surface of the object as 
\bea
a \simeq \frac{\beta}{M_P R} \frac{\Delta C}{\frac{\Delta R}{R}},
\eea
where $\Delta R=R-R_c$ is the thickness of the shell and $\Delta C$ is the variation of the chameleon field inside the shell\footnote{All the formulas mentioned above in this section are discussed in more detail in reference \cite{Zanzi:2013yy}.}.

Let us gather some useful formulas for the chameleon inside the Sun.
As we will show explicitly, if we choose $R_c \simeq \frac{9 R}{10}$ (where $R$ is the solar radius),
$M=M_P=10^{19} GeV$ everywhere constant and
$\beta \simeq n \simeq \lambda \simeq 1$,
 we obtain a chameleonic acceleration near the surface of the Sun that turns out to be roughly similar to the metric-induced acceleration near the surface of the Sun. %The same choice of parameters leads us to a similar result for the Earth: the chameleon-induced acceleration near the surface of the Earth is roughly similar to the metric-induced acceleration near the surface. 
Now to the calculations. We call $C$ the chameleon field. As already pointed out in \cite{Zanzi:2013yy}, there are models where the chameleon field is basically constant inside the shell. This result can be obtained removing the fine-tuning on the scale of the potential. Hence, let us choose $M=M_p=10^{19}$ GeV. Moreover, we choose $\beta\simeq n \simeq \lambda \simeq 1$.
If we define $\epsilon \equiv  \Delta R/R$ and $x\equiv r-R_c$, we can write \ref{formuletta} as 
\bea
\Delta C= C-C_c \simeq \frac{\beta}{M_p} \rho_c R_c^2 [\frac{x^2 }{2R_c^2}] \simeq \frac{\beta}{M_p} \rho_c R_c^2 [\epsilon^2/2],
\eea
where we made the approximation $r \simeq R$ in the shell.
The best thing would be to evaluate $R_c$ numerically, but we will simply assume $R_c \simeq \frac{9R}{10}$. In this way we have
\bea
\mid a\mid \simeq \frac{\beta}{M_p R} \frac{\mid \Delta C\mid }{\frac{\Delta R}{R}} = \frac{\beta}{M_p R} \frac{\mid \Delta C\mid }{\epsilon} \simeq \frac{\beta}{M_p \epsilon R} \cdot \frac{\beta R_c^2 \rho_c \epsilon}{6M_p} 3 \epsilon \simeq \frac{\beta^2 R \rho_c}{6M_p^2} 3 \epsilon=\frac{kRM_p^2\beta^2}{6} 3 \epsilon,
\eea
where in the last step we introduced  $k\simeq 10^{-94}$ defined by $\rho_c=kM_p^4$. 
Now we can write (in international units):
\bea
\mid a\mid \simeq \beta^2 3 \epsilon \frac{10^{-94} \cdot  10^5 km}{10^{-70} m^2} \simeq \beta^2 3 \epsilon 10^{-19} km/m^2 \simeq \beta^2 3 m/s^2 \simeq g_{\odot},
\eea
therefore, if we choose $\beta \simeq 1$ we have (near the surface of the Sun) a chameleon-induced acceleration comparable to the metric-induced acceleration.

%A similar result can be obtained for the Earth. Indeed, near the surface of the Earth, we can write
%\bea
%\mid a \mid \simeq \frac{k R_\oplus M_p^2 \beta^2}{6}, 
%\eea
%where $k\simeq 10^{-94}$ is defined as before and $R_\oplus \simeq 10^{-2} R_\odot$.

%Hence, we obtain
%\bea
%\mid  a \mid \simeq 0.1 \cdot \beta^2 m/s^2 \simeq g_\oplus.
%\eea

\subsection{Model 1}

The first possibility we explore is a model inspired to a standard chameleonic scenario but without fine-tuning on the scale of the potential. We choose a large and constant Planck mass and we fix the matter density to a reasonable value by-hand.
In particular, in the model of section \ref{approx}, we choose
$R_c \simeq \frac{9 R}{10}$, where $R$ is the solar radius.
$M=M_P=10^{19} GeV$ everywhere constant. 
$\beta \simeq n \simeq \lambda \simeq 1$.

 With our choice of parameters the chameleon is basically constant everywhere inside the Sun. One approximation in the chameleonic literature is to consider a constant value of the chameleon for $0<r<R_c$ (see for example \cite{Zanzi:2013yy}). This means that we are writing the scalar field inside the Sun as the sum of (see \cite{Zanzi:2010rs, Zanzi:2012ha} for further details) (1) a constant background value and (2) a fluctuating component. In other words we exploit a background field gauge (see for example section 17.4 of reference \cite{Weinberg:1996kr}) with a constant background field. We infer that the chameleon-mediated force is vanishing when averaged over the sphere $0<r<R_c$ because the constant background field has vanishing gradient. The fluctuating component is still present, it mediates a force on short\footnote{Short with respect to the solar radius.} distances because in the UV the chameleon has non-vanishing gradient. Hence, an interesting solar model can be constructed requiring a vanishing modification of gravity in the region $0<r<R_c$. As far as the thin-shell is concerned, we exploit the calculations of section \ref{approx} and, consequently, we modify gravity in the shell by substituting $G_N \longrightarrow 2G_N$.  We stress once again that inside the sphere $0<r<R_c$, gravity is modified at short distances: the fluctuating component of the chameleon is the relevant one in the UV, it is not constant, it has a non-vanishing gradient and, hence, it mediates a force. The force is vanishing only when averaged in the sphere $0<r<R_c$, because the background field is constant.

\subsection{Model 2}

Our second example is the standard chameleon, where we fine-tune the scale of the potential to the Dark Energy scale ($M \simeq 10^{-3} eV$) and we choose a large Planck mass $M_p \simeq 10^{19} GeV$. Moreover we choose $\beta \simeq n \simeq \lambda \simeq 1$. Interestingly, the formulas shown for model 1 do not depend on the scale of the potential $M$ and, consequently, if we assume that the thin-shell is roughly 1/10 of the solar radius, we obtain the same modification of gravity of model 1, namely, $G_N \longrightarrow 2G_N$ inside the shell. Needless to say, the scale of the potential is related to the mass of the chameleon.

\section{Solar models and chameleons} \label{smodels}

In this section we will exploit the possibility of using the sun 
as a sort of 'laboratory' for testing the existence of a  chameleon field. 
We will see how the introduction of a modification of gravity 
in a  outer shell of the sun (as  in model 1 and model 2 discussed above) alters significantly
some solar properties. Note that elsewhere in the literature, the sun, or 
other stellar structures, have been used for testing the possible existence of a new physics,
see e.g.\cite{Bottino:2002pd, Cassisi:2000hy}.

Let us first summarize some useful concepts in solar physics.

A Standard Solar Model (SSM) can be defined as a description of the solar structure providing 
(within uncertainties) the observed properties of the sun and exploiting 
initial chemical-physical characteristics 
chosen in a proper range \cite{Degl'Innocenti:1996ev}.
Usually, in these models magnetic fields and rotation are neglected and, consequently, the structure is spherically symmetric.
Under these conditions, the physical/chemical structure of the sun 
and its evolution in time  is  obtained solving a set of fundamental equations: 
mass conservation law, hydrostatic equilibrium, energy production, energy transport, equation of state and time variation
of elemental abundances due to nuclear burning and microscopic diffusion. In addition,  
for solving this set of equations  one has to know other important functions:
the solar opacity $\kappa(\rho, T)$ 
(the product $\kappa  \rho$ represents the inverse of the photon mean free path in the stellar interior)
and the nuclear cross sections,  
see e.g \cite{Stix:1991aa} for a detailed description of stellar equilibrium equations.
In particular in our calculations, 
the numerical solution of the equations mentioned above is obtained by using the stellar evolutionary code 
FRANEC \cite{Chieffo:1989aa, Ciacio:1996mk}, by adopting Livermore 2006 equation
of state (EOS) \cite{Rogers:1996kx}, Livermore radiative opacity  tables
 \cite{Iglesias:1996bh},  molecular opacities from ref. \cite{Ferguson:2005pu},
conductive opacities from ref. \cite{Cassisi:2007ty} 
%all  calculated for AS09 photospheric composition \cite{as09};
and nuclear reaction rates from NACRE compilation \cite{Angulo:1999zz}, taking
into account the astrophysical factors
$S_{1,14}, S_{3,4}$ and $S_{1,7}$\cite{Marta:2008gg, Costantini:2008ub, Junghans:2003bd} and the $^7{\rm Be}$
electron capture rate from \cite{Adelberger:1998qm}.

So in order to build a SSM one studies the evolution of an initially homogeneous solar mass model
up to the sun age ( $t_{\odot}=4.57 \; {\rm Gyr}$ \cite{Bahcall:1995bt}) so as to reproduce the main solar observables: 
luminosity $L_{\odot}= 3.8418 \cdot 10^{33}$erg/s \cite{Bahcall:2004aq}, 
radius $R_{\odot}= 6.9598 \cdot 10^{10}$cm \cite{Allen:1976aa} 
and  photospheric chemical composition, usually indicated with
the ratio $(Z/X)_{\rm ph}$ where $Z$ indicates the (relative) abundance in mass of all the elements
heavier than helium, usually called metals, and $X$ indicates the (relative) abundance in mass of hydrogen.
We adopted the most recent determination of photospheric abundances \cite{Asplund:2009fu}, corresponding to 
$(Z/X)_{\rm ph}=0.0182$. 

To achieve as  outputs the three solar observable mentioned above, one can play with  three input parameters:
%\begin{itemize}
%\item 
1) the initial abundance of helium ($Y_0$), i.e. the helium abundance at the zero age of the sun. 
The solar luminosity basically depends on $Y_0$: the smaller $Y_0$, the smaller the temperature, 
the smaller the luminosity for fixed solar age.
%\item 
2) The initial abundance of heavy  elements ($Z_0$). This quantity essentially determines the present metal content in the 
photosphere, the relationship being due to the efficiency of diffusion mechanism, since
metals are not burnt in the sun. Furthermore $Z_0$  affects
the solar luminosity since  the opacity of a star is strongly influenced by the presence of heavy elements, 
see e.g. \cite{Stix:1991aa}.
%Since $Z/X_{ph}$ is known from observational data, $Y_0$ and $Z_0$ are {\it not} independent: 
%the larger $Y_0$, the smaller the metallicity. 
3) The so called mixing-length parameter ($\alpha$). 
The radius is related to convection which dominates the energy transfer in the outer layers of the sun. 
The smaller $\alpha$, the larger the local gradient of temperature, see e.g. \cite{Stix:1991aa} for a detailed description of the
mixing length theory. 
We infer that a smaller temperature on the surface is obtained and, 
since $L_\odot $ is fixed, a bigger radius is necessary granted that we approximate the solar spectrum with a black-body one.
%\item

%\end{itemize}

Now it is crucial to observe that helioseismology has added other solar observables which have to be
reproduced:  the photospheric helium abundance $Y_{ph}$, the depth of the solar convective zone $R_b$  and 
the  sound speed profile in the solar interior.  
These quantities have been accurately determined in the last years
by helioseismic measurements (see e.g. \cite{Basu:2004zg, Degl'Innocenti:1996ev} and references therein)
and are crucial in testing solar model calculations (see e.g. \cite{Dziembowski:1998nb, Degl'Innocenti:1997cw}).
Indeed the  three  input parameters cannot be considered  free anymore.

\begin{figure}[t]
\begin{center}
\includegraphics[width=8.0 cm,height=6.5 cm,angle=0]{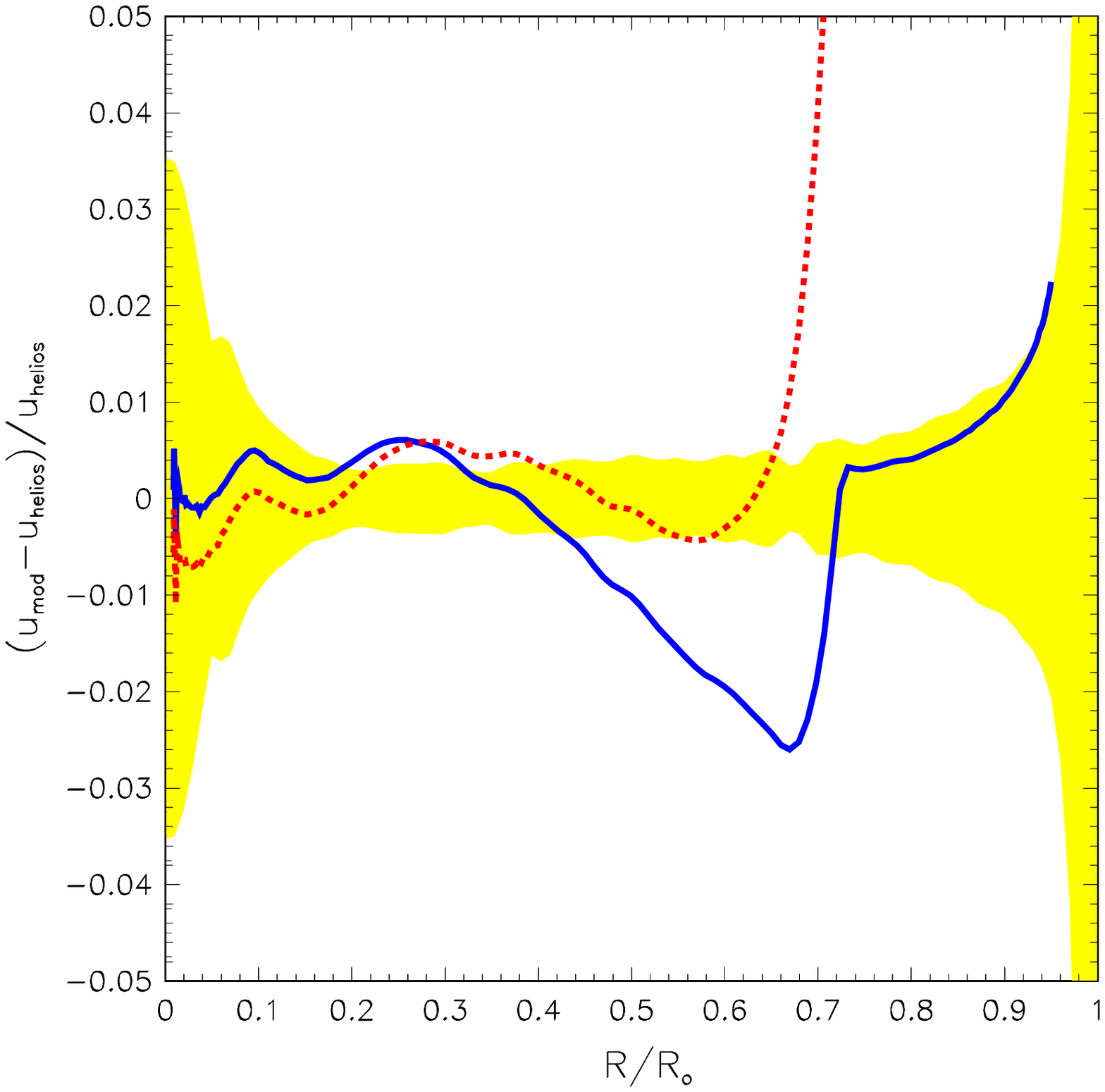}
\includegraphics[width=8.0 cm,height=6.5 cm,angle=0]{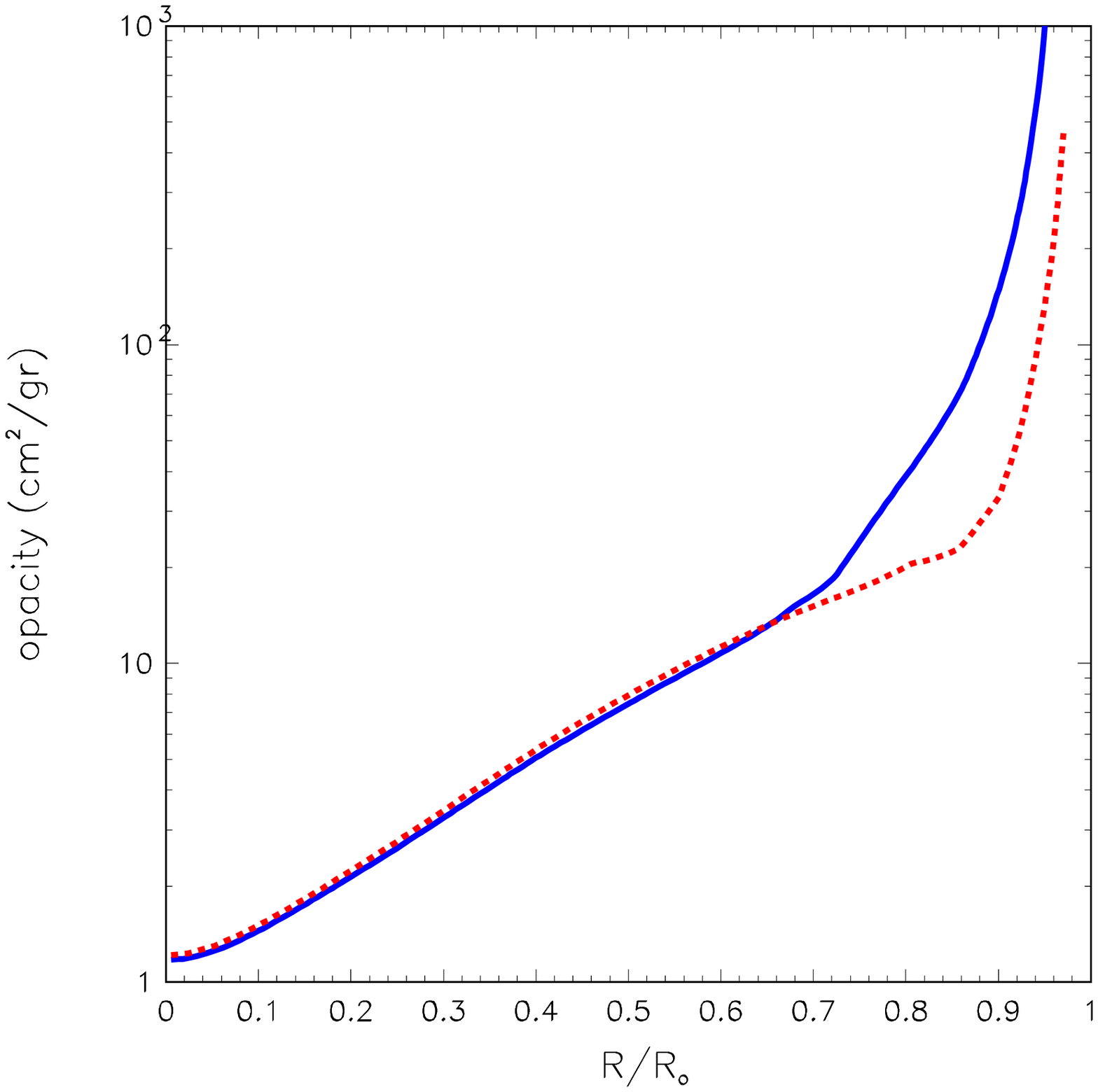}
\end{center}
\par
\vspace{-20mm} \caption{Upper panel. We report the relative difference of the
squared isothermal sound speed between our models and helioseismic predictions: $(U_{mod}-U_{helios})/U_{helios}$, for 
SSM  (continuous blu line) and non-SSM with a thin-shell of thickness $0.1 R_{\odot}$(dashed red line). The yellow shaded area represents the
conservative $3\sigma$ helioseismic uncertainties on squared isothermal sound speed as derived in \cite{Degl'Innocenti:1996ev}. 
Lower panel. We show the behaviour of  opacity along the solar profile
for our solar models, same notations as in upper panel.}
\label{figu}
\end{figure}

In Table \ref{tab1} and Fig. \ref{figu}  we report the main results of our SSM, 
obtained with all the ingredients mentioned above in this section. 
We also report the
helioseismic determinations with their more conservative uncertainties 
as derived in \cite{Degl'Innocenti:1996ev}. In particular in Fig. \ref{figu} (left panel) we report the relative difference 
of the ''squared'' isothermal sound speed ($U=p/\rho$) with respect to the helioseismic determination.
One can easily see that our SSM does not recover the helioseimic data. This is a well known and still open problem
in solar physics: standard solar models including most recent data on phothospheric composition \cite{Asplund:2009fu}
do not recover helioseismic observations,
whereas adopting older composition \cite{Grevesse:1998bj} some agreement is achieved
(see e.g. \cite{Villante:2013mba}).

%%%%%%%%%%%%%%%%%  FIGURA 1 %%%%%%%%%%%%%%%%%%%%%%

%%%%%%%%%%%%%%%%%%%%%%%%%%%%%%%%%%%%%%%%%%%%%%%%%%%5

Now let us see what occurs in the solar interior when the chameleon effect is taken into account.

We have built a non-SSM in which we have  modified artificially the gravity 
in the external solar shell, i.e.  $G_N \longrightarrow 2G_N$  for  $R > 0.9 R_{\odot}$. The results of such model
are reported again in Table \ref{tab1} and Fig. \ref{figu}. One can immediately see that the existence 
of a chameleon field in the external solar shell does not improve the agreement with helioseismology at all.

Let us discuss in some details our non-SSM model, taking into account that the aim of this section is not 
a detailed astrophysical discussion about solar modeling, but only the suggestion of 
how the chameleon physics can be phenomenologically studied in the solar interior.

In order to reproduce, at the solar age, the main solar observables (luminosity, radius, photospheric abundances)
we have varied the input parameters in our non-SSM, see Table 1. In particular, the effect of 'extra' gravity in the 
outer part of the sun determined a more efficient gravitational settling at the solar surface, 
so that in order to recover the observed
photospheric $Z/X$ one has to start with a higher $Z_0$. Let us observe that the initial metal content clearly
influences the present inner values, they are substantially coincident apart for a 10\% variation
due to elemental diffusion. Consequently with  the only increasing of $Z_0$ 
we would have an increasing of the solar opacity in the interior and consequently (at fixed age) a less bright sun. 
To compensate this effect we have to increase the initial helium abundance $Y_0$.

\begin{figure}[t]
\begin{center}
\includegraphics[width=8.cm,height=6.5 cm, angle=0]{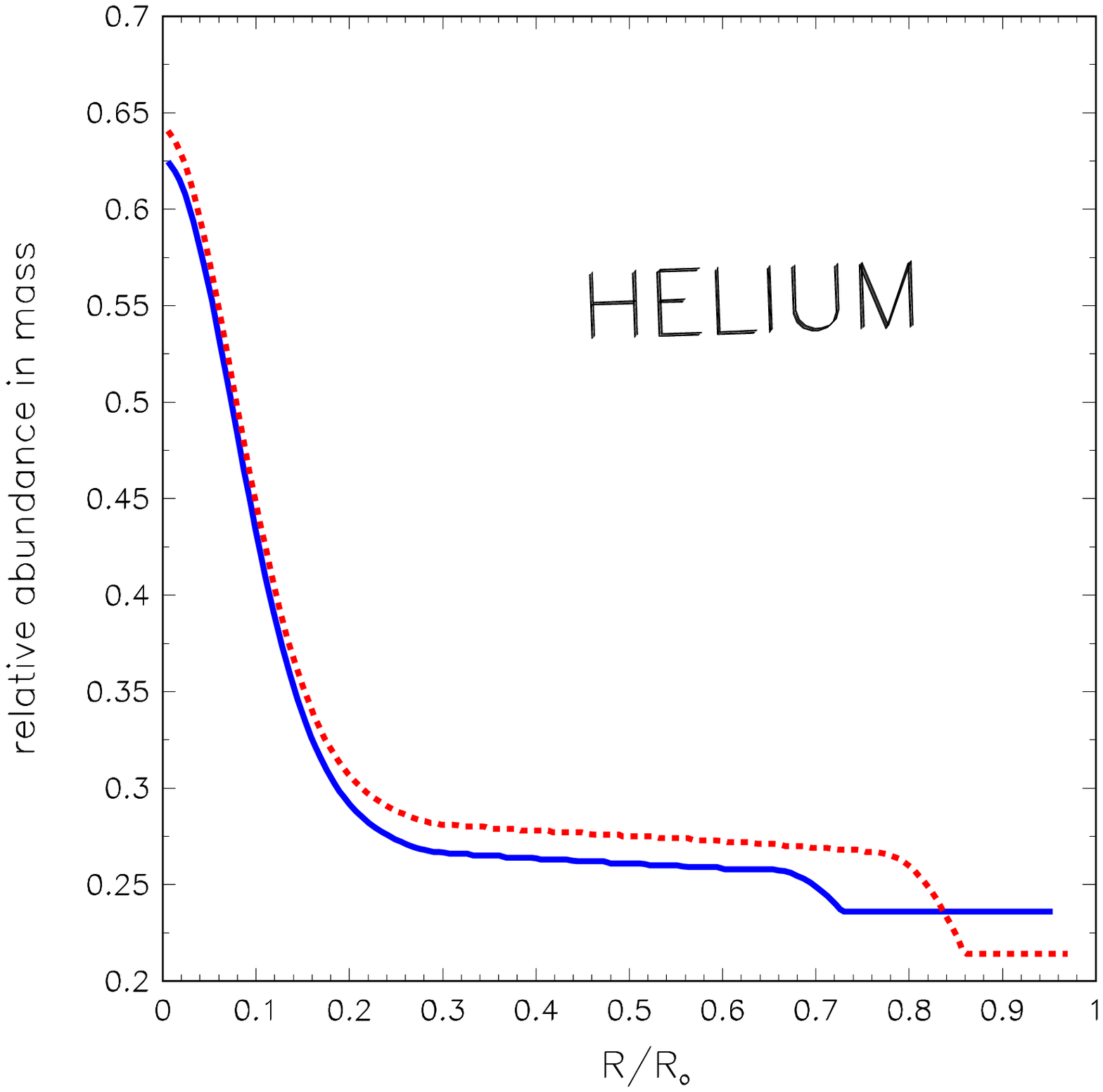}
\includegraphics[width=8.cm, height=6.5 cm,angle=0]{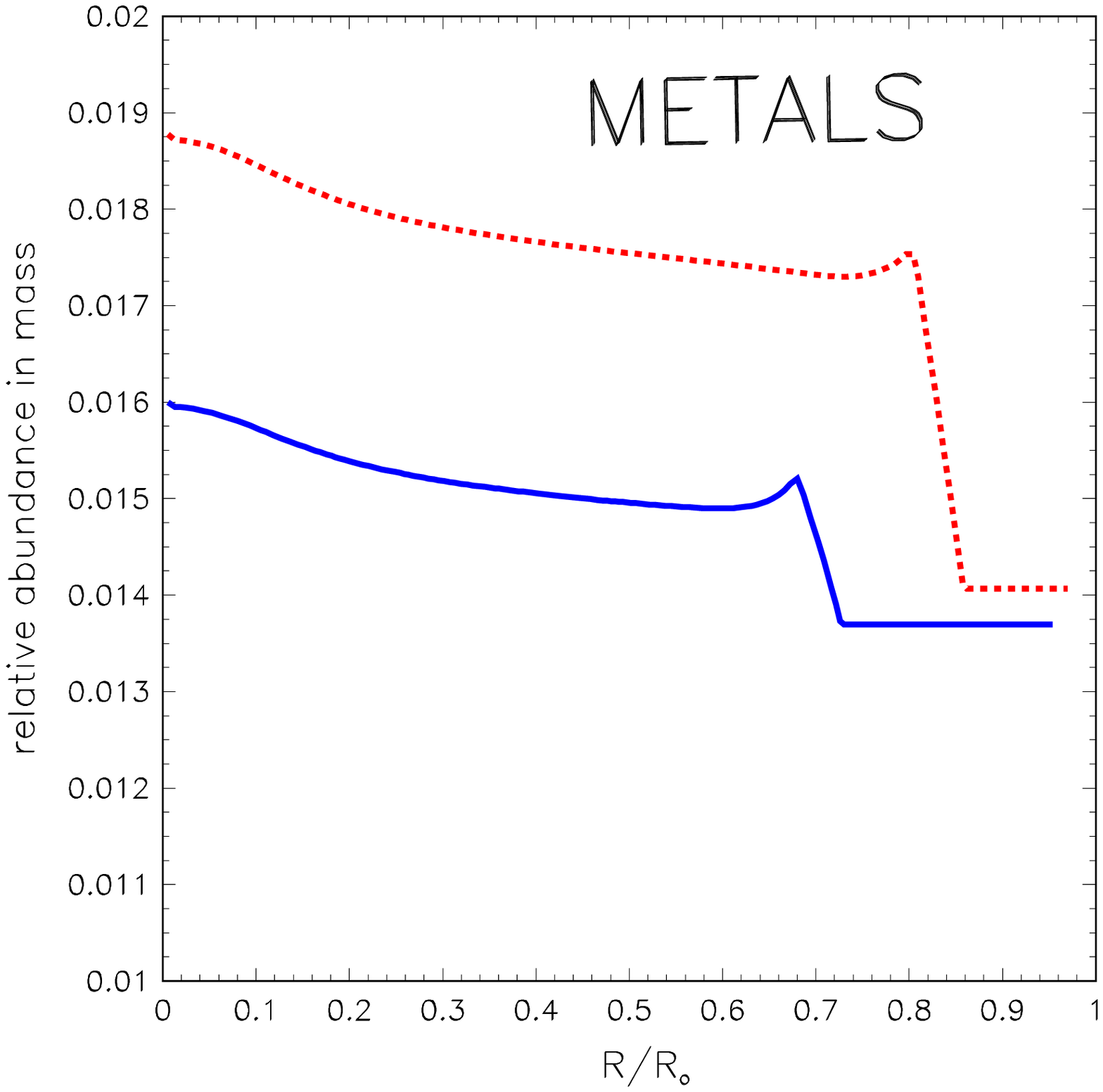}
\end{center}
\par
\vspace{-20mm} 
\caption{Upper panel: the profile of relative abundance in mass of helium along the solar structure, both for 
standard solar model SSM (continuous blue line) and non standard solar model (dashed red line). Lower panel: 
 relative abundance in mass for the metals, i.e. the elements heavier than helium, with the same notation as in upper panel.}
\label{figelements}
\end{figure}

But, even if we have increased the initial helium abundance, one sees from Table \ref{tab1} that
the present photospheric helium abundance results lower than the
values obtained in the SSM.  The gravity modification occurring
in a external region alters in such a way  the equilibrium condition of the stellar structure,
resulting in significant variations of physical and chemical properties of the sun well below 0.9$R_\odot$. 
In particular we observe that the diffusion efficiency of the elements in the region around 0.7 - 0.8 $R_\odot$ 
is increased: in Fig. \ref{figelements} we report the helium and metals profiles along the solar interior and we can see 
that the 'gap' in abundance between the radiative and the convective is almost doubled 
when we pass from SSM to non-SSM model.

Concerning the depth the convective zone $R_b$, in the solar model with modified gravity
the transition between radiative regime and convective one 
occurs in more external region.  The substantial 
modification of the elemental abundances in this part of the sun determines a decreasing of the 
solar opacity in this region. In Fig. \ref{figu} we show the profile of the solar opacity along the solar structure,
both for SSM and non-SSM model. Since in non-SSM the opacity 
is decreased for solar radius greater than $0.7R_{\odot} $,  the energy transport by radiation
remains efficient for a larger part of the sun.

Finally we observe that the squared sound speed profile is completely different with respect to the helioseismic observations,
when we are above 0.7$R_{\odot}$. Around $0.6 R_{\odot}$ the agreement is clearly better, also with respect to the SSM case, 
 this is due to the fact that generally a  less opaque sun  is preferred by helioseismic measurements
of the sound speed, see e.g.\cite{Villante:2013mba}.

In conlusion we can say that the non-standard solar model with modified gravity in an external shell, 
is in disagreement with helioseismic observations. One can observe that
we  start from a standard solar model  already in disagreement, but on the other hand we can observe that 
if we were able to build a standard solar model perfectly consistent with the central value of 
helioseismic data, the introduction of  a 'chameleon effect' would induce a  strong variation 
on $Y_{ph}$, $R_b$ and sound speed profile 
%(-10\% and +16\% respectively), 
well above the existing  ''conservative'' uncertainties  on such helioseismic observables.
%(1.3\%  and 0.14\% respectively).

%%%%%%%%%%%%%%%%%%%%%%%% FIGURA 2  %%%%%%%%%%%%%%%%%%%%%%%%%%%%

As we have seen, if we consider a standard chameleonic scenario, the modification of gravity inside the Sun induced by the presence of a chameleon field is {\it not} compatible with helioseismology. In particular, we considered two different models, namely ''Model 2'' (a standard chameleonic scenario where the Planck mass is constant and the scale of the potential is fine-tuned to the meV scale) and ''Model 1'' (a chameleonic scenario where the Planck mass is constant but the fine-tuning on the scale of the potential is absent). These two models are both inconsistent with helioseismology.

As we already mentioned above, we considered a thin-shell with $\frac{\Delta R}{R}=0.1$. The careful reader may wonder what are the theoretical grounds supporting this particular thickness and what are the phenomenological constraints on it. The evaluation of the thickness of the shell in the general case requires a numerical calculation (see \cite{Waterhouse:2006wv}), however, an estimate can be obtained from the formula (see for example \cite{Weltman:2008ll}):

\bea
\frac{\Delta R}{R} = \frac{C_{\infty}-C_c}{6 \beta M_p \Phi_c}
\eea
where $\Phi_c=\frac{M}{8 \pi M_p^2 R}$ is the Newtonian potential at the surface of the object and $C_{\infty}$ is the value of the chameleon field in the background outside the Sun. 
Remarkably, Lunar Laser Ranging constrains the difference in the free-fall acceleration of the Moon and the Earth towards the Sun to be less than approximately one part in $10^{13}$ (see for example \cite{Weltman:2008ll}) and, consequently, the bound

\bea
\frac{\Delta R_\oplus }{R_\oplus} < 10^{-7}
\eea 

is obtained. This translates into a bound for the solar thin-shell given by

\bea
\frac{\Delta R_\odot  }{R_\odot} \simeq \frac{\Delta R_\oplus }{R_\oplus} \frac{\Phi_\oplus}{\Phi_\odot} \simeq 10^{-3} \frac{\Delta R_\oplus }{R_\oplus} < 10^{-10}.
\eea

\begin{figure}[t]
\begin{center}
\includegraphics[width=8.cm,height=6.5 cm, angle=0]{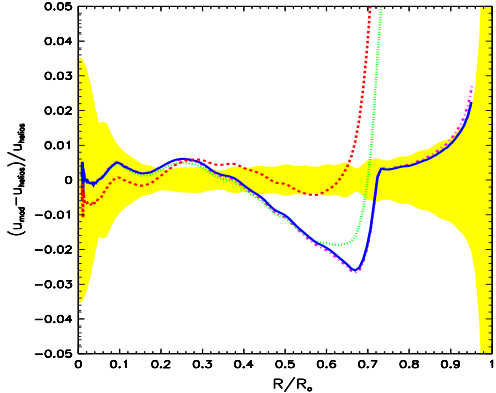}
\end{center}
\par
\vspace{-20mm} 
\caption{We report the relative difference of the
squared isothermal sound speed between our models and helioseismic predictions: $(U_{mod}-U_{helios})/U_{helios}$, for 
SSM  (continuous blu line) and non-SSMs. The dashed red line corresponds to $\frac{\Delta R}{R}=0.1$. The dashed green line corresponds to $\frac{\Delta R}{R}=10^{-4}$. The pink line corresponds to $\frac{\Delta R}{R}=2 \cdot10^{-6}$. The yellow shaded area represents the
conservative $3\sigma$ helioseismic uncertainties on squared isothermal sound speed as derived in \cite{Degl'Innocenti:1996ev}.  }
\label{modellinuovi}
\end{figure}

For this reason, we developed also solar models where the thin-shell is much smaller than $0.1 R_\odot$. The results are shown in Fig. \ref{modellinuovi} and table 1. The dashed red line corresponds to $\frac{\Delta R}{R}=0.1$. The dashed green line corresponds to $\frac{\Delta R}{R}=10^{-4}$. The pink line corresponds to $\frac{\Delta R}{R}=2 \cdot10^{-6}$ and it is the thinnest shell that can be analyzed by our numerical simulations. One can see that solar models with a shell thinner than $2 \cdot 10^{-6} R_{\odot}$ are, at the moment, basically indistinguishable from SSM. At this stage, LLR provides a constraint stronger than the one coming from solar models. 

%%%%%%%%%%%%%%%%%%%%%%%%%%  TABLE 1%%%%%%%%%%%%%%%%%%%%%
\begin{table}[t]
\center
\begin{tabular}{| c|c | c|c| c|c|}
\hline
             & SSM  & $10^{-1} R_{\odot}$  & $10^{-4} R_{\odot}$ & $2\cdot 10^{-6} R_{\odot}$ &  helioseismic \\
\hline

$Y_{0}$     & 0.268 & 0.282   & 0.269 & 0.267 & \\
$Z_0$        & 0.0153  & 0.0179   & 0.0154 & 0.0152 &   \\
$\alpha$        & 1.96  & 1.76   & 1.54 & 1.55 &   \\

\hline
$Z_{ph}$        & 0.0137   &  0.0141  & 0.0137 & 0.0136&    \\
$Y_{ph}$     & 0.236 & 0.214      & 0.235 & 0.236 &  0.249 $\pm$0.010\\
$R_b$        & 0.727 & 0.857      & 0.754 & 0.727 & 0.711 $\pm$0.003  \\
\hline
\end{tabular}\\[1cm]
\caption{We report the values for: initial helium abundance $Y_0$, initial metals abundance $Z_0$,
photospheric helium (metal) abundance, $Y_{ph}$ ($Z_{ph}$),  depth of the 
convective zone, $R_b$, both for  standard solar model (SSM), and for  non standard solar models with shell thickness in the range from $10^{-1} R_{\odot}$ to $2 \cdot 10^{-6} R_{\odot}$ , see text.
We also report the helioseismic  
values for $Y_{ph}$ and $R_b$  with their conservative $3 \sigma$ uncertainties as derived in \cite{Degl'Innocenti:1996ev}}. 
\label{tab1}
\end{table}

%%%%%%%%%%%%%%%%%%%%%%%%%%%%%%%%%%%%%%%%%%%%%%%%%%%%%%%%%%%%%%%%%%

\setcounter{equation}{0}
\section{The modified Fujii's model (MFM)}
\label{model}

In the remaining part of this article, we will further explore this connection between chameleons and solar physics. However, we will leave the standard chameleonic scenario and we will discuss a different set-up that we call ''Modified Fujii's Model'' (MFM). Indeed the MFM is the result of a lagrangian discussed by Fujii in \cite{Fujii:2002sb, Fujii:2003pa} and recently modified by one of us to discuss the cosmological constant problem in \cite{Zanzi:2010rs}. One remark is necessary. The lagrangian of the MFM is written in the string frame and, once a conformal transformation to the E-frame is performed, we obtain a chameleonic competition between a run-away potential and the $F_{\mu\nu}F^{\mu\nu}$ term of gluons \cite{Zanzi:2013xx}. In other words, in \ref{veff} the minimum in the effective potential is generated by $\rho_m$, but in the MFM, once we move to the E-frame, the minimum is formed through the presence of the $F_{\mu\nu} F^{\mu\nu}$ term of gluons with reasonable assumptions about the couplings \cite{Zanzi:2013xx} (the interactions must be a decreasing function of energy and this is natural because the IR region corresponds to the weak coupling of the string).

\subsection{The lagrangian}

Here is the model. We have two different conformal frames. On the one hand, we have the string frame (S-frame) where the cosmological constant (CC) is large (for example planckian) and the fields are stabilized (including the dilaton $\phi$). On the other hand, there is the Einstein frame (E-frame), where the dilaton $\sigma$ is a chameleon field. Remarkably, in the E-frame the amount of scale invariance is parametrized by $\sigma$: scale invariance is abundantly broken locally (i.e. ''in this room'') and it is almost restored globally (i.e. on cosmological distances). Consequently, the CC is around the meV-scale in this frame but the vacuum energy is large locally.

We write the string frame lagrangian as (the gauge part is not written explicitly but it is present in the theory)
\beq {\cal L}={\cal L}_{SI} + {\cal L}_{SB}, \label{Ltotale}\eeq where the
Scale-Invariant part of the Lagrangian is given by:

\begin{equation}
{\cal L}_{\rm SI}=\sqrt{-g}\left( \half \xi\phi^2 R -
    \half\epsilon g^{\mu\nu}\partial_{\mu}\phi\partial_{\nu}\phi -\half g^{\mu\nu}\partial_\mu\Phi \partial_\nu\Phi
    - \frac{1}{4} f \phi^2\Phi^2 - \frac{\lambda_{\Phi}}{4!} \Phi^4
    \right).
\label{bsl1-96}
\end{equation}
$R$ is the curvature. $\Phi$ is a scalar field representative of matter fields,
$\epsilon=-1$, $\left( 6+\epsilon\xi^{-1} \right)\equiv
\zeta^{-2}\simeq 1$, $f<0$ and $\lambda_{\Phi}>0$.
We can also write terms of the form $\phi^3 \Phi$, $\phi \Phi^3$ and
$\phi^4$ which are multiplied by dimensionless couplings. However
we will not include these terms. Indeed, the first two terms can be removed exploiting symmetries of strong interaction \cite{Fujii:2003pa} and the $\phi^4$ term does not clash with the solution to the CC problem, because the renormalized Planck mass in the IR region is an exponentially decreasing function of $\sigma$ (see also \cite{Zanzi:2012du}).

The Symmetry Breaking Lagrangian
${\cal L_{SB}}$ is supposed to contain scale-non-invariant terms,
in particular, a stabilizing (stringy) potential for $\phi$ in the
S-frame. For this reason we write: \beq {\cal L}_{\rm
SB}=-\sqrt{-g} (a \phi^2 + b + c \frac{1}{\phi^2}). \label{SB}
\eeq

${\cal L_{SB}}$ is a stabilizing contribution. The reader can find a possible choice of parameters in \cite{Zanzi:2012du}. For a discussion of the E-frame lagrangian and of the different dynamical behaviour of the dilaton in the two frames see \cite{Zanzi:2010rs}. Interestingly, the masses of the matter particles in the E-frame are determined by the Planck mass \cite{Zanzi:2010rs}.

As pointed out in \cite{Zanzi:2013xx}, this model satisfies a Chameleonic Equivalence Postulate (CEP).
Here is the CEP:
{\it for each pair of vacua V1 and V2 allowed by the theory there is a conformal transformation that connects them and such that the mass of matter fields $m_0,_{V1}$ (i.e. $m_0$ evaluated in V1) is mapped to $m_0,_{V2}$ (i.e. $m_0$ evaluated in V2). When a conformal transformation connects two vacua with a different amount of conformal symmetry, an additional term (in the form of a conformal anomaly) must be included in the field equations and this additional term is equivalent to the gravitational field.}

In reference \cite{Zanzi:2013xx} this has been presented as an equivalence postulate for quantum gravity. One aspect of the CEP will be particularly useful in this paper: {\it all} the ground states of the model (including the E-frame vacua) can be mapped into each other by a conformal transformation. 
We will come back to this issue in the next section.

\setcounter{equation}{0}
\section{Helioseismology}
\label{helio}

In this section we will start summarizing the hydrodynamical equations of the classical theory of solar oscillations and then we will proceed with a reinterpretation of the solar oscillations through quantum vacuum fluctuations (i.e. a quantum portrait).

Here is the classical theory.
A hydrodynamical system is specified by some functions of position {\bf r} and time $t$: the speed ${\bf v}({\bf r},t)$, the density $\rho ({\bf r},t)$ and the pressure $p ({\bf r},t)$. Mass conservation is given by the continuity equation:
 \beq
\frac{\partial \rho}{\partial t} + \nabla\cdot (\rho{\bf v})=0.
 \label{cont}
 \eeq
 If we neglect the viscosity of the gas, the relevant interactions are due to pressure and gravity and, hence, we can write the equation of motion as:

 \beq
 \rho (\frac{\partial{\bf v}}{\partial t} + {\bf v}\cdot \nabla{\bf v})=-
 \nabla p +\rho {\bf g},
 \label{moto}
 \eeq
 
where ${\bf g}$ is the gravitational acceleration, connected to the potential $\Phi$ by the definition ${\bf g}=- \nabla \Phi$.
$\Phi$ satisfies the Poisson equation

 \beq
 \Delta \Phi=4 \pi G \rho.
 \label{poisson}
 \eeq
Our last equation will be the first principle of thermodynamics $\delta q=
dE + p dV$.

Now we suggest a different interpretation of the solar oscillations from the standpoint of the E-frame of the MFM: a quantum picture.
Let us start with some preliminary remarks. In this paper we assume that the CEP is satisfied by the MFM's lagrangian and we discuss some of its consequences. It is not our intention to explore in this paper the theoretical grounds supporting this assumption (this has been already done in \cite{Zanzi:2013xx} and related references). We are well aware of the fact that the ideas we are going to present in this section cannot be considered as a rigorous proved statement, however, we think it's useful to mention this quantum picture of helioseismology as a guideline towards future research developments. 

 As already mentioned above, the CEP is telling us that, in the MFM, {\it all} the ground states (including the E-frame vacua) can be mapped into each other by a conformal transformation. We follow \cite{Zanzi:2010rs} and we suppose to live in the E-frame. Let us map the ground state of a composite hadron of the E-frame into a macroscopic ground state in the E-frame. When we perform this conformal transformation the size of the hadron is stretched to macroscopic length scales, but also the de Broglie wavelength is rescaled to macroscopic size by the same conformal factor. We infer that quantum physics is stretched to macroscopic length scales. Let us exploit these considerations to map the ground state of a composite hadron into   the ground state of the Sun. This point must be further analyzed.

As far as the hadronic radius is concerned, we can exploit the standard QCD formula

\beq 
\hbar c \simeq 200 MeV fm
\eeq
connecting the hadronic radius to $\Lambda_{QCD}$. The CEP is telling us that the same physics can be stretched to macroscopic distances. Let us explore the consequences of this statement in the case of the Sun.
Solar matter is globally a gauge singlet under $SU(3)_C \times U(1)_{em}$ and, hence, we choose a sterile spinor field to represent solar matter. Needless to say, we cannot neglect gravity and the masses of particles, therefore, the spinor field is massive. 
If we exploit the CEP, the standard QCD condensation formula is mapped into

\beq
\hbar c \simeq \Lambda_\odot R_\odot,
\eeq
where $R_\odot$ is the solar radius and $\Lambda_\odot$ is the condensation scale.

Now, what is the value of the condensation scale $\Lambda_\odot$?
The mass of the spinor field is much larger than the planckian value of the particles' masses in the Sun, because, in harmony with the CEP, we assume the spinor to be a {\it composite} field representative of {\it many} solar particles (analogously to a nucleon inside the hadron) and consequently $\Lambda_\odot>>\Lambda_{QCD}$.
Moreover, $R_\odot >> 1fm$. Hence the solar $\hbar c$ is extremely large with respect to the hadronic value. We infer that $\hbar$ is an increasing function of $\sigma$\footnote{Large values of $\sigma$ correspond to the IR region (see \cite{Zanzi:2010rs}).} and this is in harmony with the fact that quantum physics is ''stretched'' to solar length scales. Happily this peculiar $\hbar$ is fully compatible with a chameleonic behaviour of the de Broglie wavelength already discussed in this model in \cite{Zanzi:2012ha}.

This is a good point to discuss more carefully the theoretical grounds supporting quantum helioseismology. The reader may wonder whether the CEP is the only motivation for quantum helioseismology. We point out that the terms of the form $\Phi^2$ and $\Phi^4$ in the lagrangian \ref{bsl1-96} are present in the Ginzburg-Landau approach to superconductivity and they can be obtained with a mean field approximation starting from the microscopic theory (see for example \cite{Fetter:1971ll}). In other words, the structure of the lagrangian of the MFM is supporting fermion condensation in a natural way and, therefore, the description of the Sun as a macroscopic fermionic condensate, namely the quantum helioseismology portrait, is compatible with \ref{bsl1-96} (after a proper conformal transformation to the E-frame). 

To proceed further, let us analyze helioseismology from the standpoint of our quantum portrait of the Sun.
The number density of solar particles is a decreasing function of $\sigma$ because also the solar spinor field is rescaled after the conformal transformation to the E-frame. Hence, large $\sigma$ corresponds to small number density of solar particles and hence to a large solar radius. We infer that solar oscillations are related to $\sigma$-fluctuations. The variation of the solar radius is very small (with respect to the radius) and we suggest to interpret these oscillations as small quantum fluctuations of the dilaton around its minimum. 

Interestingly, the model of \ref{Ltotale} is motivated, at least partially, by string theory \cite{Zanzi:2012bf} and the string length is chameleonic in this model \cite{Zanzi:2012ha}. We infer that the strings near the solar surface might be macroscopic and quantum in this model: the oscillations of the solar radius, in this model, might be simply macroscopic (and quantum!) string vibrations. An interesting line of development will try to explore potential phenomenological consequences of this comment on higher-spin fields \cite{Sagnotti:2003qa} (remarkably when the matter density is small the string mass is small and the massive spectrum of the string is lowered to smaller energies).

We are led to a reinterpretation of the forces which govern the dynamics of the Sun. Classical helioseismology is telling us that the solar radius is stabilized by the competition between gravity and pressure (which is the result of electromagnetic interaction). Interestingly, in quantum helioseismology, the stabilization of the chameleonic dilaton (i.e. of the solar radius) is the result of a competition between gravity, which is attractive, and the gradient of the vacuum energy, which gives a repulsive contribution to matter. Since pressure creates the correct competition with gravity in the classical picture, the two contributions (pressure and gravity) must have a similar energy scale already at the classical level. This result is automatic in the quantum picture, because pressure is related to the (gradient of the) planckian vacuum energy and its value is fixed by the conformal anomaly which is equivalent to gravitation \cite{Zanzi:2013xx}. In classical helioseismology the oscillations near the surface of the Sun are maintained by pressure (the so-called p-modes) and, hence, in the MFM, the solar oscillations near the surface are a direct result of a variation of the chameleonic vacuum energy. In other words, the role of the {\it electromagnetic} interaction in classical helioseismology is played, in quantum helioseismology, by the variation of the quantum vacuum energy and this vacuum energy is determined, in our model, by the {\it gravitational} interaction. The length scale of quantum fluctuations is related to the mass of the field and, therefore, in quantum helioseismology the length scale of solar fluctuations is related to the mass of the chameleon.

One interesting lesson from quantum helioseismology is that the QCD dynamics  is mapped, through a conformal transformation, to a (quantum) gravitational dynamics governing solar physics. In other words, it seems at this stage that, in the MFM, (quantum) gravitation is the IR limit of QCD.  In this sense, there might be no difference in this model between strong interaction and gravity: they might be just two aspects of the same interaction. In a modern language, it seems that the CEP might support a unification of all the interactions (a natural result in heterotic string theory) and the strength of this interaction is simply parametrized by the value of $\sigma$. Further research efforts are necessary to support properly this idea. 

Some words of caution are necessary. First of all, the stringy origin of the lagrangian \ref{Ltotale} is still under investigation, the connection to string theory is only partial at the moment. Moreover, more detailed and quantitative motivations for introducing quantum helioseismology would be welcome and many aspects of this proposal are still waiting for a careful investigation. For these reasons, quantum helioseismology can be considered a conjecture at this stage, even if we assume to describe Nature through our chameleonic model (where the CEP is satisfied).

\section{Conclusions}
\label{conclusions}

In this article we discussed some aspects of solar physics from the standpoint of the so-called chameleon fields (i.e. quantum fields, typically scalar, where the mass is an increasing function of the matter density of the environment). Here is a summary of our results.\\ 1) We analyzed the effects of a chameleon-induced deviation from standard gravity just below the surface of the Sun. In particular, we developed solar models which take into account the presence of the chameleon and we showed that they are inconsistent with helioseismic observations. This inconsistency presents itself not only with the typical chameleon set-up discussed in the literature (where the mass scale of the potential is fine-tuned to the meV), but also if we remove the fine-tuning on the scale of the potential. However, if we modify standard gravity only in a shell of thickness $10^{-6} R_{\odot}$ just below the solar surface, the model is basically indistinguishable from a Standard Solar Model. At this stage lunar laser ranging provides a stronger constraint than the one coming from solar models. \\ 2) We pointed out that, in a model recently considered in \cite{Zanzi:2010rs}, a conceivable interpretation of the solar oscillations is given by quantum vacuum fluctuations. In this model, called MFM, helioseismology is quantum physics. In particular, we assumed that the CEP is satisfied by the MFM's lagrangian and we discussed some of its consequences. It was not our intention to explore in this paper the theoretical grounds supporting this assumption (this has been already done in \cite{Zanzi:2013xx} and related references). We are well aware of the fact that these ideas cannot be considered as a rigorous proved statement, however, we think it's useful to illustrate this quantum picture of helioseismology as a guideline towards future research developments.

Interestingly, the lagrangian of the MFM can be obtained, at least partially, from string theory. If the connection with the string was complete, we could consider quantum helioseismology as a step forward in string phenomenology. Needless to say, we live in the LHC era, but we think that it is important to search for alternative ways of creating links between particle physics and experiments.

%%%%%%%%%%%%%%%%%%%%%%%%%%%%%%%%%%%%%%%%%%%%%%%%%%%%%%%%%%%%%%%%%%%%%%%%%%%%%%%%%%%%%%%%%%%%%%%%%%%%%%%%%%%%%%%%%%%%%
% ACKNOWLEDGEMENTS
%%%%%%%%%%%%%%%%%%%%%%%%%%%%%%%%%%%%%%%%%%%%%%%%%%%%%%%%%%%%%%%%%%%%%%%%%%%%%%%%%%%%%%%%%%%%%%%%%%%%%%%%%%%%%%%%%%%%%
%\vspace{0.5cm}
\subsection*{Acknowledgements}

Special thanks are due to Giovanni Fiorentini and Franco Strocchi for useful comments and suggestions. B.R. is extremely grateful to Scilla Degl'Innocenti and to Pisa Astroparticle group for useful comments and for providing the updated version of opacity tables. We also acknowledge the referees of the journal for pointing out useful comments about a first edition of the manuscript.

%%%%%%%%%%%%%%%%%%%%%%%%%%%%%%%%%%%%%%%%%%%%%%%%%%%%%%%%%%%%%%

%%%%%%%%%%%%%%%%%%%%%%%%%%%%%%%%%%%%%%%%%%%%%%%%%%%%%%%%%%%%%%%%%%%%%%%%%%%%%%%%%%%%%%%%%%%%%%%%%%%%%%%%%%%%%%%%%%%%%
% BIBLIOGRAPHY
%%%%%%%%%%%%%%%%%%%%%%%%%%%%%%%%%%%%%%%%%%%%%%%%%%%%%%%%%%%%%%%%%%%%%%%%%%%%%%%%%%%%%%%%%%%%%%%%%%%%%%%%%%%%%%%%%%%%%

\providecommand{\href}[2]{#2}\begingroup\raggedright\endgroup

\end{document}